\documentclass[]{article}
\usepackage{emulateapj}
\usepackage{epsfig}
\usepackage{ifthen}
\submitted{draft}

\submitted{Submitted to the Astrophysical Journal}

\def\gsim{\;\rlap{\lower 2.5pt
 \hbox{$\sim$}}\raise 1.5pt\hbox{$>$}\;}
\def\lsim{\;\rlap{\lower 2.5pt
   \hbox{$\sim$}}\raise 1.5pt\hbox{$<$}\;}
\def\msol{{\rm\,M_\odot}}

\def\kpc{{\rm\,kpc}}

\def\kms{{\rm\,km\,s^{-1}}}

\def\spose#1{\hbox to 0pt{#1\hss}}
\def\lta{\mathrel{\spose{\lower 3pt\hbox{$\mathchar''218$}}
     \raise 2.0pt\hbox{$\mathchar''13C$}}}
\def\gta{\mathrel{\spose{\lower 3pt\hbox{$\mathchar''218$}}
     \raise 2.0pt\hbox{$\mathchar''13E$}}}

\def\zsol{{\,Z_\odot}}

\begin{document}
	
\title{Metal Enrichment of the Intergalactic Medium at $z=3$ by
Galactic Winds}

\author{Anthony Aguirre,\footnote{Institute for Advanced Study, School of Natural Sciences, Princeton NJ 08540}$^{,b}$
Lars Hernquist,\footnote{Department of Astronomy, Harvard University
60 Garden Street, Cambridge, MA 02138}
Joop Schaye,$^{a}$
David H. Weinberg,\footnote{Department of Astronomy, Ohio State University, Columbus, OH 43210}
Neal Katz,\footnote{Department of Astronomy, University of Massachusetts, Amherst, MA 01003}
\& Jeffrey Gardner\footnote{Department of Astronomy, University of Washington, Seattle, WA 98195}
}
\setcounter{footnote}{0}

\begin{abstract}

	Studies of quasar absorption lines reveal that the low density
intergalactic medium at $z\sim3$ is enriched to $10^{-3}-10^{-2}$
solar metallicity.  This enrichment may have occurred in an early
generation of Population III stars at redshift $z\ga 10$, by
protogalaxies at $6 \la z \la 10$, or by larger galaxies at $3 \la z
\la 6$.  This paper addresses the third possibility by calculating the
enrichment of the IGM at $z \ga 3$ by galaxies of baryonic mass $\ga
10^{8.5}\msol$. We use already completed cosmological simulations to
which we add a prescription for chemical evolution and metal ejection
by winds, assuming that the winds have properties similar to those
observed in local starbursts and Lyman-break galaxies.  Results are
given for a number of representative models, and we also examine the
properties of the galaxies responsible for the enrichment as well as
the physical effects responsible for wind escape and propagation.  We
find that winds of velocity $\ga 200-300\kms$ are capable of enriching
the IGM to the mean level observed, though many low-density regions
would remain metal free.  Calibrated by observations of Lyman-break
galaxies, our calculations suggest that most galaxies at $z \ga 3$
should drive winds that can escape and propagate to large radii.  The
primary effect limiting the enrichment of low-density IG gas in our
scenario is then the travel time from high- to low-density regions,
implying that the metallicity of low-density gas is a strong function
of redshift.

\end{abstract}

\section{Introduction}

The detailed comparison of quasar absorption spectra with the
predictions of cosmological hydrodynamic simulations has established
that the Ly$\alpha$ absorption `forest' is caused by a
smoothly-fluctuating neutral component of the intergalactic medium
(IGM) (e.g., Cen et al. 1994; Zhang, Anninos, \& Norman 1995;
Hernquist et al. 1996).  The simulations also reveal a strong
correlation between absorber H\,I column density and gas overdensity
$\delta\equiv \rho_{\rm gas}/\langle\rho_{\rm gas}\rangle$, so that
the study of $N({\rm H\,I})\la 10^{15}{\rm \,cm^{-2}}$ absorbers gives
information about the IGM for $\delta \la 10$ at $z \ga 3$ (e.g. Zhang
et al. 1998; Dav\'e et al. 1999).  Furthermore, studies of C\,IV,
Si\,IV and O\,VI have revealed that the low-density IGM is not -- as was
long expected -- chemically pristine, but has been enriched to $\sim
10^{-2.5}\zsol$ (with a scatter of $\sim1\,$dex) down to $N(H\,I)
\approx 10^{14.5}$ ($\delta \sim 5$) (e.g., Meyer \& York 1987;
Songaila \& Cowie 1996), and perhaps to even lower overdensities
(e.g., Cowie \& Songaila 1998; Ellison et al. 2000; Schaye et
al. 2000).

This enrichment may have occurred during a Population III
star-formation epoch at $z \gg 5$ (e.g. Carr, Bond \& Arnett 1984;
Ostriker \& Gnedin 1996; Haiman \& Loeb 1997; Abel et al. 1998), later
through the disruption of small `protogalaxies' at $6-8 \la z \la
10-20$ (e.g. Gnedin \& Ostriker 1997; Gnedin 1998; Madau, Ferrara \&
Rees 2001), or even more recently by the ejection of metals from
fairly massive galaxies at $z \la 6-8$.  Earlier enrichment might
allow more uniform IGM pollution (depending on how strongly biased the
star formation is), but is constrained by the amount of metal that can
be generated at early times; conversely, sufficient metal is easily
available at late times, but the larger inter-galaxy spacing may
preclude a homogeneous metal distribution (which observations may or
may not indicate).

In this paper, we address enrichment of the IGM by supernova-driven
winds from galaxies of (gas+star) masses $\ga 10^{8.5}$ at $z \la 8$.
Our calculations employ a smoothed-particle hydrodynamics (SPH)
simulation with $128^3$ dark matter particles and $128^3$ SPH
particles in a $(17\,{\rm Mpc})^3$ box, adopting a cosmology with
$\Omega_\Lambda=0.6$, $\Omega_b=0.047$ and $\Omega_m = \Omega_{\rm
CDM}+\Omega_b = 0.4$. The simulation is described in more detail in
Aguirre et al. (2001) and in Weinberg et al. (1999).  These
simulations do {\em not} include winds, but we use a parameterized
prescription, described in \S~\ref{sec-method}, to add metals in a way
that models their ejection and distribution by supernova-driven
galactic outflows.  This post-processing method allows us to easily
vary our assumptions and parameters, and we give results for a number
of representative models. We then compare the resulting
intergalactic metal distribution to the quasar absorption line
observations and discuss the implications for wind enrichment of the IGM.

\section{Method}
\label{sec-method}

The method by which we calculate IGM enrichment is discussed in 
detail in Aguirre et al. (2001).  Briefly, our method employs a limited
number of outputs from already-completed SPH simulations that include
star formation.  Each unit of forming stellar mass instantaneously
generates $y_*$ units of metal mass ($y_*=\zsol$ is assumed for all
models).  We then deposit this metal mass in nearby gas particles as
follows:
\begin{enumerate}
\item{Particles are grouped\footnote
{The grouping is done using the SKID package, publicly available at
http://www-hpcc.astro.washington.edu/tools.} into bound objects (`galaxies').
The mass of
star formation is tallied for a given galaxy, for which we also
compute the mass, star-formation rate (SFR), and an area $A = 44(M/10^{11}\msol)^{0.76}\,{\rm kpc^{2}}$ based on
an empirical relation between disk area and mass (Gavazzi, Peirini \&
Boselli 1996).}
\item{Galaxies with SFR/A $>$ SFR$_{\rm crit} = 0.1\msol{\rm
yr^{-1}\,kpc^{-2}}$ are assumed to coherently drive steady-state, mass-conserving
winds with velocity $v_{\rm out}$ at the center of star formation and
with a mass outflow rate of $\dot m = 1.3\chi(600{\,\rm
km\,s^{-1}}/v_{\rm out})^2$ times the SFR.  Here $\chi$ parameterizes
the fraction of supernova kinetic energy in the wind, if we assume
$10^{51}\,$ergs are released per 100$\msol$ of star
formation.  Based on observations that generally $\dot m \sim 1 \msol{\rm yr^{-1}}$ 
(Heckman et al. 2000), 
we use $\chi=1$.
We allow galaxies with SFR/A $<$ SFR$_{\rm crit}$
to drive winds also, but with the wind energy attenuated by 
$[({\rm SFR/A})/(\rm SFR_{\rm crit})]^4$.}
\item{We divide the volume about the center of star formation into
$N_a$ regions of equal solid angle about directions
$(\theta_i,\phi_i)$, with $\sim 16$ galactic gas particles per region
(thus $N_a$ increases with galaxy mass).}
\item{For each angle $(\theta_i,\phi_i)$ we integrate the equations of
motion of a `test shell' beginning at radius $r_0$ containing 10\% of
the galaxy's mass.  The shell at radius $r$ feels accelerations due to
gravity, the ram pressure of the wind (attenuated by the potential
difference between $r$ and $r_0$), the ambient thermal pressure, and
the sweeping up of a fraction $\epsilon_{\rm ent}$ of the ambient
medium (matter could be left behind by the shell either if the ambient
medium is inhomogeneous, or as the shell fragments and leaves part of
itself with lower velocity.  We take a fiducial value of
$\epsilon_{\rm ent}=0.1$). All accelerations are computed using the
simulation particles contained in the solid angle about
$(\theta_i,\phi_i)$ (see Aguirre et al. 2001 for details).
Eventually, the shell stalls (i.e. has a negligible velocity with
respect to the local medium) at a radius $r_{\rm stall}(\theta,\phi)$.
We also stop the shell if/when its propagation time exceeds the time
between its launch and the redshift at which we quote our results.}
\item{In each angular region, metals are distributed among gas
particles within $r_{\rm stall}(\theta,\phi)$ such that the metal mass in a shell at
$r$ of thickness $dr$ is $\propto (r/r_{\rm stall})^\alpha\,dr.$ We
generally take $\alpha=3$, but the resulting enrichment is rather insensitive
to $\alpha$ for $2 \la \alpha \la 5$ (Aguirre et al. 2001).}
\item{If $r_{\rm stall} < 2r_0$ for some angle, the corresponding metal mass
is not distributed as per the wind prescription, but is deposited `locally'
over the 32 gas particles nearest to its progenitor star, 
using the SPH smoothing kernel (see, e.g., 
Hernquist \& Katz 1989).}
\end{enumerate}
The process is repeated for each galaxy, at each time step.  New stars
are formed with the metallicity of the gas from which they form.
Metals accumulate in gas particles 
(which move between time steps),
and the distribution of metals in stars and in galactic and
intergalactic gas can be assessed at each time step.

\section{Results}

\subsection{Enrichment of the IGM}
We first give results for six illustrative models.  The first four
have all their parameters fixed (to the `fiducial' values given above)
except for the wind outflow velocity $v_{\rm out}$, which we take to be
$600\pm200\kms$, $300\pm200\kms$, $200\pm100\kms$, or $100\pm50\kms$
(uniformly distributed in the given range for each model), independent of
galaxy mass (as suggested by Heckman et al. 2000).  Next, we
fix $v_{\rm out}=300\pm200\kms$ and vary the entrainment fraction $\epsilon_{\rm ent}$
from 1\% to 100\%.

	The key results of the calculations are shown in
Fig.~\ref{fig-windres}.  The top two panels give a sparse sampling of
individual particle metallicities, versus the gas overdensity
$\delta$. The stellar yield $y_*$ is uncertain by perhaps a factor of
two, and all of the curves could be scaled vertically for a higher
assumed value.\footnote{The yield can be constrained using the
metallicities of observed galaxies; see Aguirre et al. (2001).}  The
metallicity at $\delta \la 10^4$ could also be (roughly) scaled down
by a factor $(1-Y_{\rm ret})$ if a fraction $Y_{\rm ret}$ of the metal
produced is assumed not to be incorporated into the wind.  Panel~A
shows that in the $600\kms$ model, the enrichment is fairly uniform.
A significant fraction of particles (shown in a bar at the bottom with
an artificial 1 dex of scatter added) have zero metallicity, but the
rest have a metallicity spread at a given overdensity of $\sim$1-2
dex.  In contrast, the $300\kms$ model (panel B) has {\em some} metal
rich particles at low densities, but many more pristine regions that
the winds have not reached.

Panels C and D show the median and mean gas metallicities
vs. overdensity for all six models.  The shaded box indicates the
rough range of metallicity found in most absorption systems for which
metal lines are detected.  The area with vertical lines indicates the
region where the metallicity is agreed upon by independent groups (e.g., 
Rauch, Haehnelt, \& Steinmetzet 1997; Dav\'e et al. 1998).
The horizontal lines indicate yet smaller overdensities to which this
metallicity may extend (e.g., Ellison et al. 2000; Schaye et
al. 2000).  Even if the observational results were complete, neither
the mean nor the median particle metallicities we quote are directly
comparable to the observational results, but should bracket the
quantity that is comparable.  The mean would be too high because
(especially in cases with inhomogeneous distributions) it tends to be
dominated by a small number of high-metallicity regions; the median
gives information about the {\em number} of lines that would be metal
rich, but is an underestimate because an absorption line would
generally probe more than one simulation gas particle, decreasing the
scatter.

 \vbox{ \centerline{ \epsfig{file=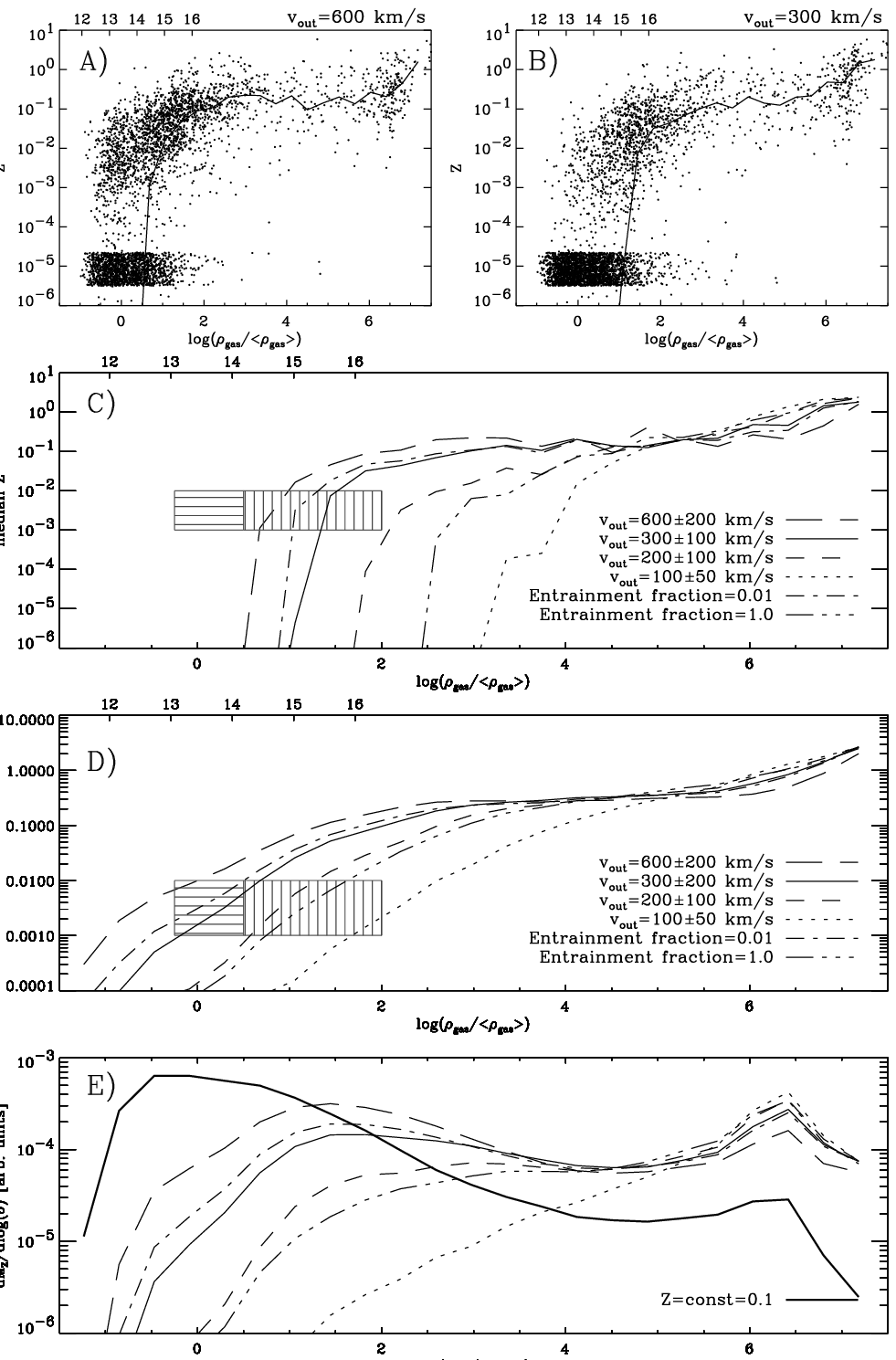,width=9.0truecm}}
\figcaption[]{ \footnotesize Enrichment of the IGM at $z=3$ as a
function of gas density. {\bf Panel A:} Random subsample (1 in 500) of
particle metallicities for wind model with $v_{\rm out}=600\kms$,
versus overdensity $\delta \equiv \rho_{\rm gas}/\langle\rho_{\rm
gas}\rangle$.  Top axis (here and in all panels) gives approximate
$\log N(H\,I)$, using the relation of Schaye (2001; assuming his
fiducial parameters).  The solid line shows the median metallicity
versus $\delta$. {\bf B:} As for panel A, but in a $v_{\rm
out}=300\kms$ model. {\bf C:} Median metallicity versus $\delta$ for
wind models with $v_{\rm out}=100,200,300,600\kms$, and for `minimal'
and `maximal' outflow models (as described in text).  The shaded box
roughly indicates the metallicity of low-column density Ly$\alpha$
clouds. {\bf D:} As for panel C, but {\em mean} metallicities are
plotted. {\bf E:} As for panel C, but gives mean metallicities times
the fraction of baryons at a given $\delta$, showing the contribution
by components with different $\delta$ to the cosmic metal density. The
thick solid line shows the distribution assuming constant metallicity (with
the same total metal mass).
\label{fig-windres}}}
\vspace*{0.5cm}

In panel E the gas metallicity is multiplied by the fraction
of gas at the given overdensity, showing what fraction of all cosmic
metal lies at a given density.  For the models that can plausibly
account for the observed metal lines in the IGM, $\sim 10-30\%$ of all metal is
extragalactic (EG), i.e. is not in cooled gas or stars in galaxies

The curves show that the assumed wind velocity at inception is quite
important in the resulting IGM enrichment.  Slow winds tend to be
confined, as can be seen from the small fraction ($2\%$) of all metals
outside of galaxies in the $100\kms$ model (see panel E). Those that escape
tend to stall at relatively small distances where the density is still
high, and fail to effectively pollute the low-density IGM.  Increasing
$v_{\rm out}$ both increases the total EG metal density (raising the
EG fraction to 50\% for $v_{\rm out}=600\kms$) and shifts the metals
to lower density regions (increasing the $\delta=1$ mean enrichment by
a factor of several hundred).  Changes in the entrainment fraction
also have a fairly large effect.  If (instead of the fiducial 10\%)
only 1\% of the ambient gas were entrained, 20\% of all metals would be  
outside of galaxies in our fiducial 300\,km/s model; while only 4\% would be
extragalactic if all ambient gas were entrained.  Changing our other
model parameters within a range of reasonable values has a smaller
but still substantial effect; doubling SFR$_{\rm crit}$ or changing
$\chi$ by a factor of two changes the ejection fraction by $\la 30\%$
and changes the mean enrichment at $\delta=1$ by $\la 50\%$ (Lowering
SFR$_{\rm crit}$ would have little effect since most galaxies at $z >
3$ are already driving winds for our assumed value.)

Two methodological points concerning our results merit
discussion. First, resolution tests reveal that galaxies with $\ga
6\times 10^8\msol$ in baryons are well-resolved.  Galaxies below this
limit exist (and are used in the analysis) but will be
under-represented both in their number and in their star-formation
rates (Weinberg et al. 1999).  Studies of dwarf galaxies at high $z$
have concluded that at $z \la 6$, galaxies of baryon mass $\la 5\times
10^7\msol$ are suppressed by photo-evaporation (e.g., Ferrara \&
Tolstoy 1999), leaving a (redshift-dependent) `gap' in mass where
galaxies may exist but are not resolved by our simulations.

Second, our method necessarily neglects the effect of the outflows on
the galaxies.  Galaxy formation studies with ineffective feedback
generally predict more low-mass galaxies than are observed, and the
simulations used here may over-estimate the cosmic SFR and stellar
mass for this reason (see Weinberg et al. 1999 and Aguirre et
al. 2001).  While we cannot quantify exactly the effect these two
(opposing) inaccuracies will have on our results, comparison of the
mass functions for different numerical resolutions indicates that we
can probably hope to estimate the enrichment by $z \la 6$ galaxies to
within a factor of a few.~\footnote{This assumes that the missing
low-mass galaxies are distributed like the higher-mass galaxies. If
low mass galaxies are strongly biased toward appearing in low-density
regions, then the results may differ significantly when they are included.}

Because IGM enrichment is so sensitive to $v_{\rm out}$, we can set a
lower limit on that quantity if the observed enrichment is primarily
due to winds of the type discussed in this paper.  To achieve the
{\em mean} metallicity of $10^{-2.5}\zsol$ observed in $N({\rm
H\,I})\sim 10^{14.5}$ absorbers, $v_{\rm out}\ga 200\kms$ is
necessary, and $v_{\rm out} \ga 300\kms$ would be needed to enrich
absorbers down to $N({\rm H\,I})\sim 10^{13.5}$ with a similar
metallicity (though this may {\em over}-enrich higher column density
systems).  A higher velocity is probably necessary if the observed
enrichment is fairly uniform; for a {\em median} metallicity of
$\sim10^{-2.5}\zsol$ down to $N({\rm H\,I})\sim 10^{14.5}{\rm
\,cm^{-2}}$, a wind speed of $v_{\rm out} \ga 300\kms$ is probably
required, though a more detailed analysis is necessary for a robust
comparison.  These conclusions depend on the assumed entrainment
fraction (and more weakly on the other parameters), but changing these
assumptions cannot increase the enrichment very much because
enrichment is always limited by the available time between shell
launch and $z=3.$ Thus if the observed enrichment turns out to be
uniform at column densities $N({\rm H\,I})\ll 10^{14.5}{\rm
\,cm^{-2}}$, enrichment from smaller and/or higher redshift galaxies
would be required.

\subsection{Physics of outflows}

\begin{figure*}
\vbox{ \centerline{ \epsfig{file=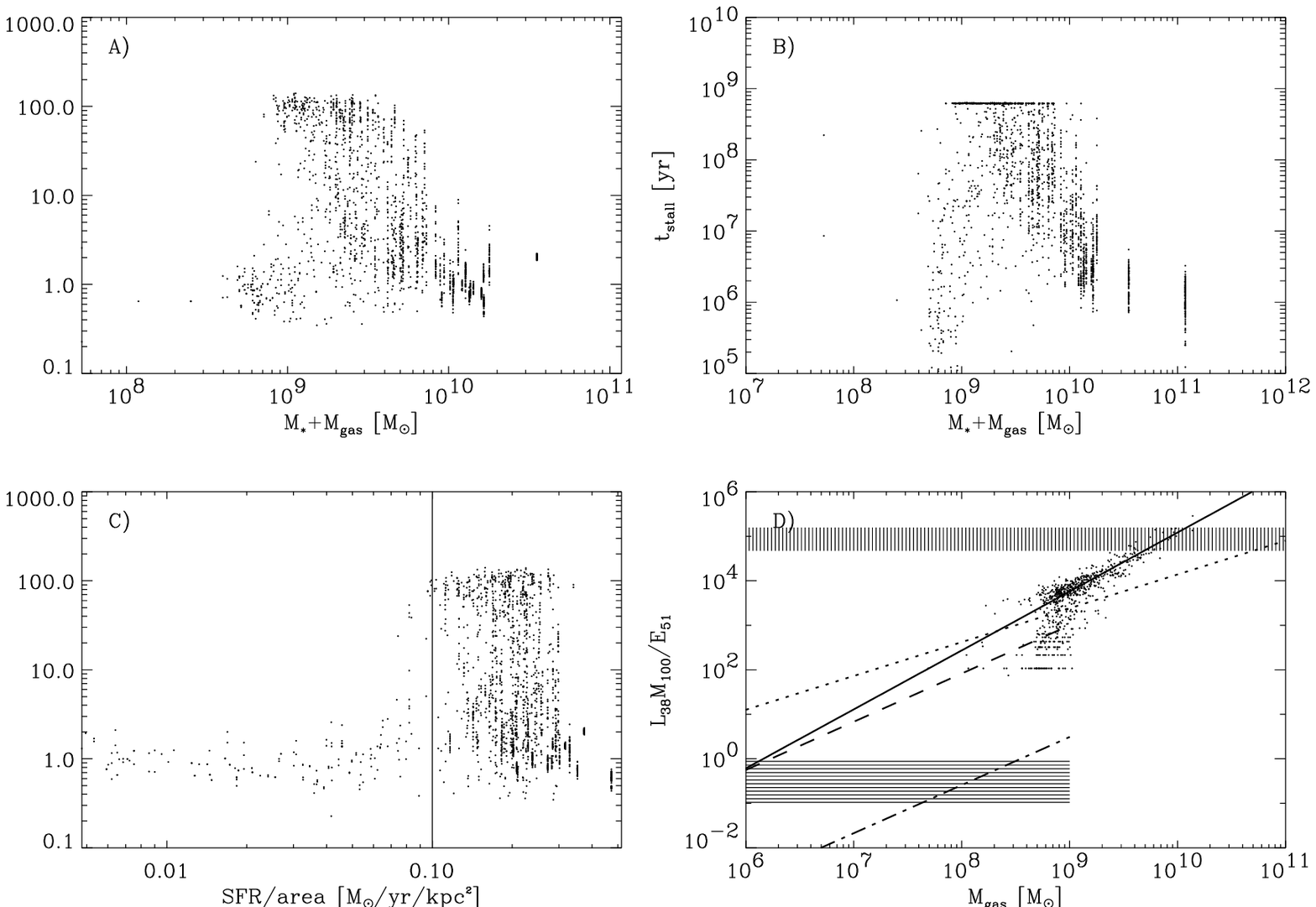,width=18.0truecm}}
\figcaption[]{ \footnotesize Quantities at $z=4$ for wind-driving
galaxies. {\bf Panel A:} Wind stopping radius $r_{\rm stall}$ (i.e. where the
shell has small velocity w.r.t. the ambient medium or runs out of time at
$z=3$) versus galactic baryonic mass.  {\bf B:} Time between shell
launch at $z=4$ and stalling; the strip just below $10^9\,$yr
corresponds to the time between $z=4$ and $z=3$.  {\bf C:} $r_{\rm
stall}$ vs. total SFR/area of galaxy.  {\bf D:} Supernova energy
generation rate in units of $10^{38}{\rm\,erg\,s^{-1}}$ versus galactic 
baryonic mass, assuming
$10^{51}E_{\rm 51}$\,ergs per supernova and 1 supernova per $100M_{\rm
100}\msol$ of star formation. Points show the simulated galaxies, to
which the solid line is fit.  The dotted line corresponds to our
fiducial value of SFR$_{\rm crit}$.  The upper (vertically) shaded
region is the range of SFRs ($10-50\msol/$yr) found by Pettini et
al. (2001) converted in the same manner.  The lower shaded region is
the parameter space probed by Mac Low \& Ferrara (1999).  The dashed
and dot-dashed lines are the critical luminosities derived by Silich
\& Tenorio-Tagle (2001) for wind escape from spherical and disk
galaxies, respectively.
\label{fig-wqz3}}}
\vspace*{0.5cm}
\end{figure*}

In addition to providing estimates of intergalactic enrichment, our
calculations also shed light on the physics of wind escape.
Figure~\ref{fig-wqz3} gives information about the outflows propagating
between $z\approx4$ and $z=3$ from individual galaxies in the $v_{\rm
out}=300\kms$ model.  Panel A shows the shell stalling radius for each
shell about a sample of 500 simulation galaxies; each point
corresponds to one propagation direction and each vertical `stripe' to
a galaxy. At $3 \le z \la 4$ about half of all galaxies are driving winds that
reach $r_{\rm stall} \gg 10\,$kpc.  The variation in $r_{\rm stall}$
for a given galaxy is due to anisotropies in the galaxy and IGM;
variations in $r_{\rm stall}$ among galaxies of a given mass are due
to combined variations in the outflow velocity, the SFR and the
properties of the nearby IGM.  Since we have assumed that wind
velocity is independent of galaxy mass, the most massive galaxies
drive winds ineffectively because of their relatively large potential
wells.  The cutoff at $r_{\rm stall} \sim 100\,$kpc is due to the
finite time between $z \sim 4$ and $z=3$ (where we stop each shell's
integration).  This is shown in Panel B, which displays the stalling
time for the shell segments (one point for each angle).  As shown in
Panel C, at $z=4$ almost all of the galaxies exceed the assumed
critical SFR/(area).  But as long as this threshold is satisfied, the
areal SFR does not strongly influence how well winds escape.

	Panel D allows a comparison of our results for wind escape
with two other investigations.  The points in the figure indicate the
supernova energy injection rate for each galaxy in units of
$10^{38}\,$ergs, converted from star formation rates by assuming that
one supernova releasing $10^{51}E_{\rm 51}$\,ergs forms per per
$100M_{\rm 100}\msol$ of star formation.  The vertically shaded region
indicates the range of SFRs measured in bright, wind-driving
Lyman-break galaxies by Pettini et al. (2001), converted in the same
way.  The solid line is a fit that extrapolates the simulation
results to lower-mass galaxies, and the dotted line represents our
fiducial critical SFR/area (i.e. we assume that galaxies above the
dotted line are driving winds coherently -- though the winds may
ultimately be retained as shown in panel A).  Mac Low \& Ferrara
(1999; see also Ferrara \& Tolstoy 2000) have studied galactic winds
in dwarf galaxies using numerical simulations, for the range of galaxy
masses and supernova energy injection rates shown by the shaded box.
They find that only the lowest-mass galaxies ($10^6\msol$) can
`blow-away' the bulk of their ISM, though higher mass galaxies can
allow `blowout' where hot, metal rich gas can escape.  Their results
agree roughly with the estimates of Silich \& Tenorio-Tagle (2001),
who find that winds can only escape spherical galaxies (corresponding
to `blowout') for energy injection rates above the dashed line of
Panel D; winds can escape flattened galaxies for rates above the
dot-dashed line.  According to our calculations, winds can escape the
simulation galaxies (which are not highly flattened), even for
entrainment fraction unity, from galaxies with mass $\gg
10^6\msol$. This would seem to conflict with the findings of Mac Low
\& Ferrara (1999), but the disagreement is probably due to the rather
small energy injection rates they assume for their largest galaxies --
$10^4$ 

\vbox{ \centerline{ \epsfig{file=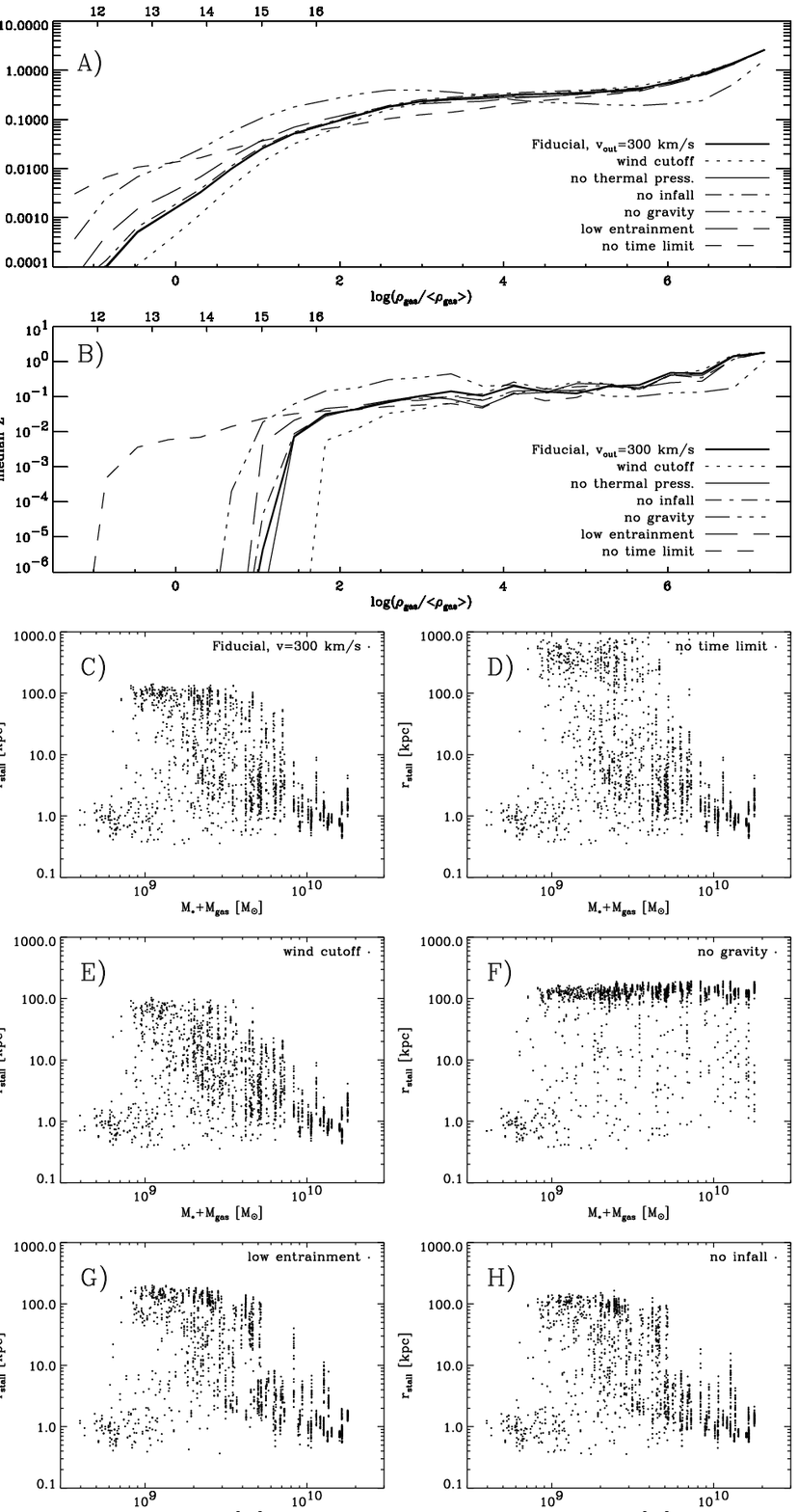,width=9.0truecm}}
\figcaption[]{ \footnotesize Enrichment of the IGM at $z=3$ for the
fiducial model with $v=300\kms$, with various effects turned off.
{\bf A:} Mean metallicities versus $\delta$ for wind models with
gravity, wind ram pressure, IGM thermal pressure, IGM infall, or the
time limit turned off.  {\bf B:} As in panel A, but median
metallicities.  {\bf C-F:} Wind stalling radius for wind propagating
between $z=4$ and $z=3$, versus galaxy baryon mass, with various
effects turned off (see text).
\label{fig-forces}}}
\vspace*{0.5cm}

\noindent times smaller than the rate indicated by the cosmological
simulations.  The simulated galaxy SFRs at high $z$ are high enough to
fulfill the (extrapolated) `blow-away' criterion of Silich \&
Tenorio-Tagle for all galaxy masses, as are the SFRs of the
Lyman-break galaxies (unless the galaxies have masses $\ga
10^{11}\msol$.)

	In Figure~\ref{fig-forces} we show plots of both IGM
enrichment and stalling radius vs. mass in the fiducial model with
each effect on shell propagation turned off in turn.  First (panel D)
we remove the time constraint.  This shows that some shell fragments
from $z=4$ galaxies would ultimately 

\vbox{ \centerline{ \epsfig{file=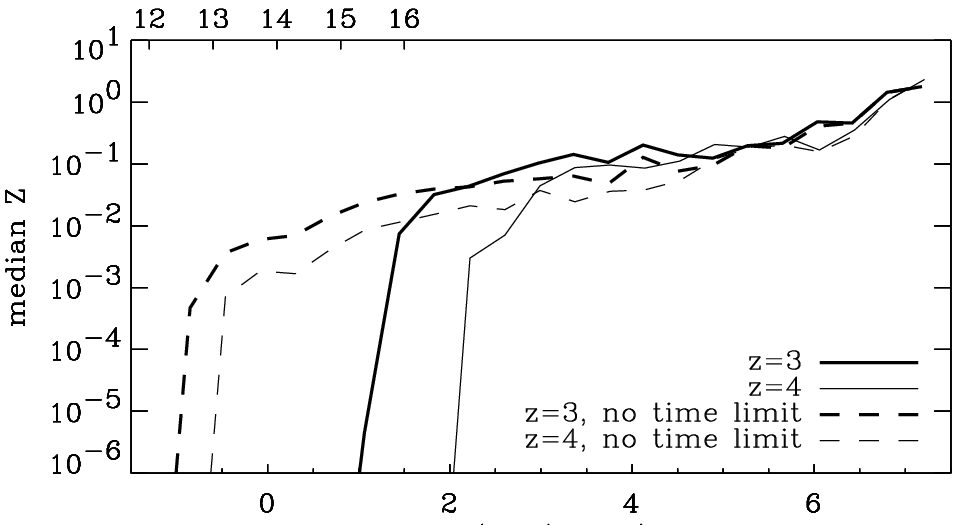,width=9.0truecm}}
\figcaption[]{ \footnotesize Median particle metallicity
vs. overdensity at $z=3$ and $z=4$, with or without wind propagation
stopped at the given redshift.
\label{fig-times}}}
\vspace*{0.5cm}

\noindent propagate to $\sim 1\,$Mpc (the
cutoff comes from the distance $d$ at which $Hd=v_{\rm out}$).  Panel
E shows the results if the 
wind is turned off $10^8\,$years
after the shell launch.  This 
is roughly the minimal interval $\tau_d$ over which a
starburst `event' could drive the shell, since the energy deposition
rate is roughly constant for $\tau_e \sim 4\times 10^7\,$yr after a
coeval starburst (Tenorio-Tagle et al. 1999), and the wind (or thermal
pressure) will still drive the shell for the time $\tau_c$ it takes the
wind (or sound) to cross the shell radius.  If the shell speed is
comparable to the wind speed, $\tau_e \sim \tau_c$ so $\tau_d \sim 2 \tau_e$.
The effect of truncating the wind is 
noticeable but slight.
In panel F we have turned off gravity, and essentially all
of the shells escape, showing that in this case gravity is the
dominant effect in confining winds.  Of the shells that still do not
escape, about half are retained due to thermal pressure, and about
half due to the sweeping up of matter.  Panel G shows the effect of
very low mass entrainment: winds escape to slightly larger radii on
average, but can actually increase the escape radius in some cases because 
higher entrainment adds more momentum to the shell.
Entrainment actually causes as much deceleration on the shell as
gravity; but since gravity also attenuates the wind (decreasing the
acceleration of the shell), it has a larger overall
effect. Panel H shows that taking into account infall from the IGM is
not very important, but tends to slightly decrease the
stalling radius. In summary, we find that only the hot wind, the
entrainment of ambient material and gravity are important in
determining shell propagation in most cases at $z=3$.  More hot gas
(especially in clusters) and stronger infall, however,
might make thermal pressure and IG infall more important at low $z$.

  The most important factor limiting the enrichment of very low
density regions is the time constraint as shown in
Fig.~\ref{fig-times}. If we assume that the winds can propagate
`instantaneously' into the IGM, they reach quite low-density regions,
and the enrichment at $z=3$ is only slightly higher and more uniform
than at $z=4$.  But if the time limit is respected, the enrichment
spreads to significantly lower-density regions between $z=4$ and $z=3$
due to the extra Gyr of propagation time.

\section{Discussion and Implications}

Metal enrichment of the $z \sim 3$ IGM has also been investigated by
Gnedin \& Ostriker (1997), Gnedin (1998) and Cen \& Ostriker (1999)
using simulations that include chemical evolution.  Gnedin finds that
supernova blow-apart of protogalaxies is ineffective, but that
dynamical removal of metals can explain the metals in low-density
regions -- though this requires a metal formation rate at $z \ga 4$
that is about 10-50 times that observationally inferred at $z\sim 3-4$
(e.g., Steidel et al. 1999) and implies more efficient dynamical
removal of metals than found by Aguirre et al. (2001).

Cen \& Ostriker (1999) assume a fixed metal ejection fraction into
regions of $\sim 400\kpc$ (the resolution of their grid-based
hydrodynamical treatment; see Cen \& Ostriker 2000).  Their results
for mean metallicity are similar to our $v=200\kms$ model, though
their metallicity distribution at fixed $\delta$ shows almost no
scatter at $z=3$ (but becomes more inhomogeneous with time).  In
contrast, our calculations find a highly inhomogeneous metallicity
distribution, which becomes somewhat more uniform with time.  This may
be due to the rather low spatial resolution of their simulations or
may indicate that stars are forming (and expelling metals into the
IGM) in very low density regions in Cen \& Ostriker's simulations, in
contrast to ours.  Comparison of quasar absorption spectra to
simulated spectra shows that $\sim 0.5$ dex of scatter in gas
metallicity (among absorbers in which metals are detected) may be
necessary to explain the variation in observed C\,IV/H\,I ratios
(Rauch et al. 1997; Hellsten et al. 1997; Dav\'e et
al. 1998), so this point merits further investigation, but will
probably require the generation of synthetic spectra from our
simulations.

	An important physical effect that is not addressed by our
treatment is that of the outflowing winds on the IGM: if metals are
transported directly from galaxies to very low density regions by
movement of large quantities of gas, then this kinetic energy must be
absorbed by the IGM. Yet the widths of low-column density Ly$\alpha$
forest lines are typically $20-30\kms$, and line ratios indicate
photoionization, not collisional ionization (e.g. Songaila \& Cowie
1996).  Our method does not allow us to assess the effects of winds on
the IGM (see Theuns, Mo \& Schaye [2000] for some discussion), but we
can make some general comments.  If expanding shells of swept-up gas
maintained their integrity and propagated to roughly the inter-galaxy
spacing, they would completely re-make the structure of the Ly$\alpha$
forest, except perhaps in deep voids.  But this is not the expected
scenario: upon leaving a galaxy, the wind and `shell' will likely mix
into a complex, multi-phase outflow consisting of fragmented shell(s),
the hot wind, dense clouds of entrained (but not swept-up) ambient
medium, etc.  Small, dense blobs might maintain fairly low temperature
through efficient cooling, and could reach large radii before stalling
and mixing with the IGM (and by definition when the wind stalls it has
small local velocity and will not greatly heat the medium).  In
addition, the hot wind itself will not be observable in absorption,
and will flow more more easily in low-density directions. Hence it may
not seriously disrupt the relatively dense filaments that give rise to
much of the Ly$\alpha$ absorption.

	These issues merit further study because our results, combined
with observations of high-$z$ galaxies, indicate that significant
outflow from $z \ga 3$ galaxies seems inevitable.  Essentially all of
the simulated galaxies at $z \ga 3$ 

\vbox{ \centerline{ \epsfig{file=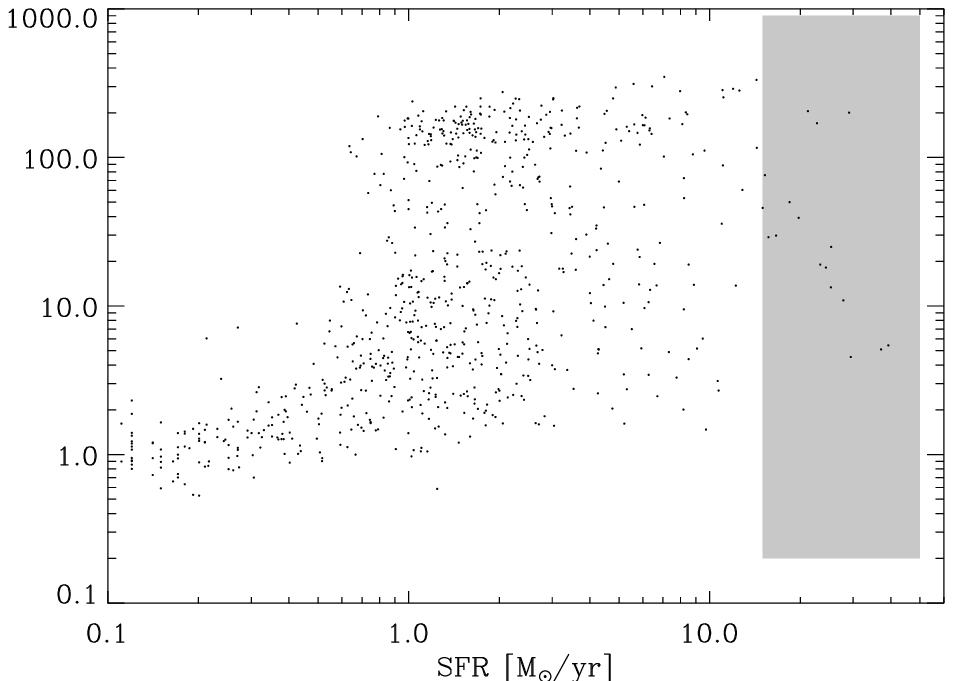,width=9.0truecm}}
\figcaption[]{ \footnotesize Maximum stalling radius in each galaxy
vs. star formation rate, at $z=4$, for $v_{\rm out} = 750\pm500\kms$
and entrainment fraction unity. The Lyman-break galaxies of Pettini et
al. (2001) show SFRs of $10-50\msol\,{\rm yr^{-1}}$.
\label{fig-lybr}}}
\vspace*{0.5cm}

\noindent display specific time-averaged
SFRs exceeding the observed threshold for driving superwinds in local
(starburst) galaxies -- and stochastic star formation should result in
much higher rates for short periods.  Moreover, available observations
of $z\sim3$ galaxies reveal that they {\em are} driving winds, with
characteristic velocities exceeding those of our fiducial model:
Pettini et al. (2001) find that Lyman break galaxies at $z\sim 3$ with
SFRs of $\sim 10-50\msol$/yr drive winds at $\sim 100-800\kms$ (as
measured by absorption lines) or $\sim 250-1250\kms$ (in Ly$\alpha$
emission).  These galaxies correspond to the highest ($\sim
10^{10}\msol$) mass simulated galaxies (see Fig.~\ref{fig-wqz3} and
Dav\'e et al. 1999).  The outflow signatures in the spectra are quite
similar to those in the nearby wind-driving starbursts upon which we
base our method, so the observed velocities should roughly correspond
to our assumed initial outflow velocities.  In this case, as shown in
Fig.~\ref{fig-lybr}, the outflows tend to escape the galaxies even for
an entrainment fraction unity, and propagate to large distances --
none of the forces we include in our calculations would be capable of
containing the outflow.  Thus we predict that winds from Lyman-break
galaxies do escape those galaxies' potentials and enrich the
IGM,\footnote{If the Lyman break galaxies actually represent
lower-mass, very strongly star-bursting galaxies this conclusion is only
strengthened, though it may then be less applicable to the rest of the
$z\sim3$ population.} and expect that this would hold for lower mass
galaxies as well.

\section{Conclusions}

	We have simulated the enrichment of the $z=3$ IGM by galactic
winds from $\ga 10^{8.5}\msol$ galaxies, assuming that most drive
winds with properties similar to those observed in local starbursts
and in Lyman-break galaxies.  We find that for winds with outflow
velocity $v_{\rm out}\ga100\kms$ enrichment is significant, and that
if $v_{\rm out}\ga 300\kms$, the winds can (roughly) account for
the metallicity of the low-column density
($N({\rm H\,I})\sim 10^{14.5}{\rm\,cm^{-2}}$) Ly$\alpha$
forest.  Higher velocities could enrich yet lower-density components
of the IGM. 

	Our treatment of wind-driven outflows includes forces due to
gravity, thermal pressure, ram pressure of the ambient medium, and the
hot wind itself.  We find that at $z \sim 3$ thermal pressure and the
the infall velocity of the IGM are generally negligible, but that
outflows can be confined by gravity or ambient material for
sufficiently high mass galaxies and/or low velocity winds.  Our
results suggest that the winds observed in Lyman break galaxies should
escape and travel to fairly large distances.  If the Lyman break
galaxies represent high-mass, non-starburst galaxies, then the
lower-mass $z=3$ galaxies should drive outflows to large distances as
well, unless outflow velocity depends strongly on galaxy mass.
The effect of such winds on the physical and thermal
structure of the IGM is essentially unknown.
The good agreement between simulations like these and the observed
statistical properties of the Ly$\alpha$ forest suggest that 
the effect cannot be extremely large.

We find that the main effect preventing the enrichment of very
low-density regions of the IGM is the time available for metal-rich
material to propagate from galaxies (which are concentrated in
filaments) into the voids.  This leads to a picture in which a given
set of galaxies enriches a progressively larger cosmic volume with
decreasing redshift, and we predict that if low-density regions are
enriched by winds from $\ga 10^{8.5}\msol$ galaxies at $z \la 6$, the
median metallicity of low-density regions should be a very strong
function of redshift.

The scenario described here, enrichment of the IGM by massive
($\ga 10^{8.5}\msol$) galaxies at modest ($z \la 6$) redshift,
is but one of several interpretations for the origin of metal
in the Ly$\alpha$ forest.  Other possibilities, distinguished by
epoch and mass scale, include enrichment by population III stars
at $z \sim 20$, or the disruption of low-mass protogalaxies at
$z \sim 10$.  How are we to decide which, if any, of these three
scenarios is most relevant to the real Universe?
 
Based on our modeling, we anticipate that these mechanisms will
predict differences in the metal distribution of the IGM that can be
tested observationally.  In addition to the mean enrichment, the
dispersion in metallicity at a given overdensity and the trend of
metallicity with overdensity can both be measured.  It seems natural
to expect that metal deposition at relatively higher redshift will
lead to a rather different distribution of metals than that inferred
from our analysis.  Thus, future modeling, combined with observational
samples of increasingly large size, may discriminate between the
various scenarios.  To the extent that this goal can be realized, the
metals discovered in the $z \sim 3$ Ly$\alpha$ forest may provide a
direct probe of the conditions responsible for the production of the
first metals in the Universe.
 
\acknowledgements

We thank T. Heckman for useful comments and information on superwinds.
This work was supported by NASA Astrophysical Theory Grants NAG5-3922,
NAG5-3820, and NAG5-3111, by NASA Long-Term Space Astrophysics Grant
NAG5-3525, and by the NSF under grants ASC93-18185, ACI96-19019, and
AST-9802568.  JG was supported by NASA Grant NGT5-50078 for the
duration of this work, and AA was supported in part by the National
Science Foundation grant no. PHY-9507695 and by a grant in aid from
the W.M. Keck Foundation.  The simulations were
performed at the San Diego Supercomputer Center.

\end{document}